\documentclass[final,1p,times]{elsarticle}

\usepackage{amssymb}
\usepackage{amsmath}
\usepackage[hidelinks]{hyperref}
\usepackage{xcolor}
\hypersetup{
	colorlinks,
	linkcolor={red!50!black},
	citecolor={blue!50!black},
	urlcolor={blue!80!black}
}

\journal{Building and Environment}

\begin{document}


\begin{frontmatter}



\title{Reverberation Time Control by Acoustic Metamaterials {in a Small Room}}


\author[inst1,inst2]{Sichao Qu}
\cortext[cor1]{Corresponding author.}

\author[inst2]{Min Yang\corref{cor1}}
\ead{min@metacoust.com}

\author[inst2]{Yunfei Xu}
\author[inst2]{Songwen Xiao}
\author[inst1]{Nicholas X. Fang}

\affiliation[inst1]{organization={Department of Mechanical Engineering, The University of Hong Kong},
            addressline={Pokfulam Road}, 
            city={Hong Kong},
            postcode={999077}, 
            country={China}}

\affiliation[inst2]{organization={Acoustic Metamaterials Group Ltd.},
            addressline={Data Technology Hub, TKO Industrial Estate}, 
            city={Hong Kong},
            postcode={999077}, 
            country={China}}

\begin{abstract}
In recent years, metamaterials have gained considerable attention as a promising material technology due to their unique properties and customizable design, distinguishing them from traditional materials. This article delves into the value of acoustic metamaterials in room acoustics, particularly in small room acoustics that poses specific challenges due to their significant cavity resonant nature. Small rooms usually exhibit an inhomogeneous frequency response spectrum, requiring higher wall absorption with specific {spectrum} to achieve a uniform acoustic environment, i.e., a constant reverberation time over a wide audible frequency band. To tackle this issue, we developed a design that simultaneously incorporates numerous subwavelength acoustic resonators at different frequencies to achieve customized broadband absorption for the walls of a specific example room. {The on-site experimental measurements agree well with the numerical predictions, attesting to the robustness of the design and method.} The proposed method of reverse-engineering metamaterials by targeting specific acoustic requirements has broad applicability and unique advantages in small confined spaces with high acoustic requirements, such as recording studios, listening rooms, {and car cabins}.
\end{abstract}



\end{frontmatter}


\section{Introduction}

Improving the auditory experience in enclosed spaces, particularly achieving a flat frequency response in a small room, has been an enduring challenge for room acoustics \cite{kuttruff2012room}. When recording or listening to natural sounds, it's important to have an acoustic environment that can reproduce the sounds accurately without distortion \cite{parker2009good}. In scenarios such as {car interior}, telephone booth and music practice room, a more uniform acoustic feedback is also comfortable for the users. However, small rooms tend to have oscillating responses due to a sparse modal distribution in frequency. Also, the sound fields as function of space can be highly inhomogeneous and anisotropic \cite{garai2016sound,nolan2020isotropy}. To improve the acoustic environment, absorbing materials can be introduced to modify the resonant response. In theory, achieving complete flattening of the system response necessitates uneven specific boundary absorption in frequency, contingent upon the system's original response spectrum and the desired acoustic requirements. While traditional porous acoustic materials \cite{sagartzazu2008review,shin2020sound} can provide effective broadband absorption, their customization flexibility across frequencies is limited by its inherent porous structures and related dissipative nature. To effectively improve the acoustic environment of a small room, new acoustic materials or structures that allow customization of absorption performance over a wide bandwidth are needed. This article explores how acoustic metamaterials, an emerging technology in recent years, can offer a novel approach and effective solution to this problem.

Based on local resonances \cite{ma2016acoustic,yang2017sound}, acoustic metamaterials have exhibited many intriguing properties \cite{cummer2016controlling,gu2021controlling,xue2022topological} (such as negative refraction, perfect absorption, and topological transport, etc.) that were previously impossible for natural materials. Unfortunately, these properties depend highly on specific modes, so the desired functions can only be achieved at a single frequency or a few discrete frequencies. However, among the properties, absorption is one of the exceptions. Many researchers have found that broadband absorption is completely feasible by combining different resonant units \cite{jimenez2017rainbow,jiang2014ultra,long2019broadband}. In 2017, Yang et al. \cite{yang2017optimal} proposed a general theoretical framework for designing broadband absorption metamaterials through the integration of Fabry–Pérot (FP) resonances, which pushed the absorption performance to the theoretical limit considering the thickness-absorption trade-off induced by causality constraint \cite{rozanov2000ultimate,acher2009fundamental,meng2022fundamental}. Another less recognized aspect is that the distribution of resonances can enable global and flexible control over the absorption spectrum of metamaterials \cite{yang2018integration,qu2022microwave}. This phenomenon is analogous to selective absorption in optics, which has various applications such as radiative cooling \cite{raman2014passive}, solar power generation \cite{li2019scalable}, and image processing \cite{wesemann2019selective}. Here, we will explore how this characteristic can be leveraged to address the challenges associated with improving the acoustic environment of small rooms \cite{ma2018shaping,qu2020minimizing,wang2022controlling}.

While the human perception of sound may be complex due to factors such as spaciousness, naturalness, and warmth, these indicators are closely associated with the reverberation time of indoor spaces  \cite{kaplanis2014perception,raer2016statistics}. As an initial study, this article aims to achieve a uniform reverberation time across different frequencies by determining the ideal absorption spectrum of a room's walls and inversely designing metamaterial absorbers. To this end, we use the Eyring equation to calculate the reverberation time above the Schroeder frequency \cite{schroeder1987statistical} and the quality factor of the resonant system to calculate the lifetime of sound energy within the room below this frequency. As an example, we consider a small room with dimensions of $1.1\times2.1\times2.1\,\mathrm{m}^3$ [Figure \ref{fig1}(b)] and design a 6.8 cm-thick metamaterial absorption unit that utilizes a Helmholtz resonator and 44 FP resonators to target a uniform reverberation time ($T_{60}$) of 0.1 seconds above 175 Hz. Finally, we verify the effectiveness of the proposed approach by installing the designed metamaterials on three walls of a real room to regulate the reverberation time. The results demonstrate that the metamaterial-based solution outperforms the traditional {foam-based} approach by a significant margin.

In what follows, we defined the small sample room's geometry and material properties in Section~\ref{sec:2}. { The method of determining the wall materials' absorption efficiency in different frequency bands according to the target reverberation time is described in detail in Section \ref{sec:3}.} In Section~\ref{sec:4}, we designed a broadband metamaterial absorber fulfilling the target and validated it through impedance tube experiments. In Section~\ref{sec:5}, we applied the large-area metamaterials in a real testing room and conducted on-site measurements.  We conclude in Section~\ref{sec:6}.

\section{Model Definitions\label{sec:2}}

\begin{figure}[b!]
	\includegraphics[width=12.0cm]{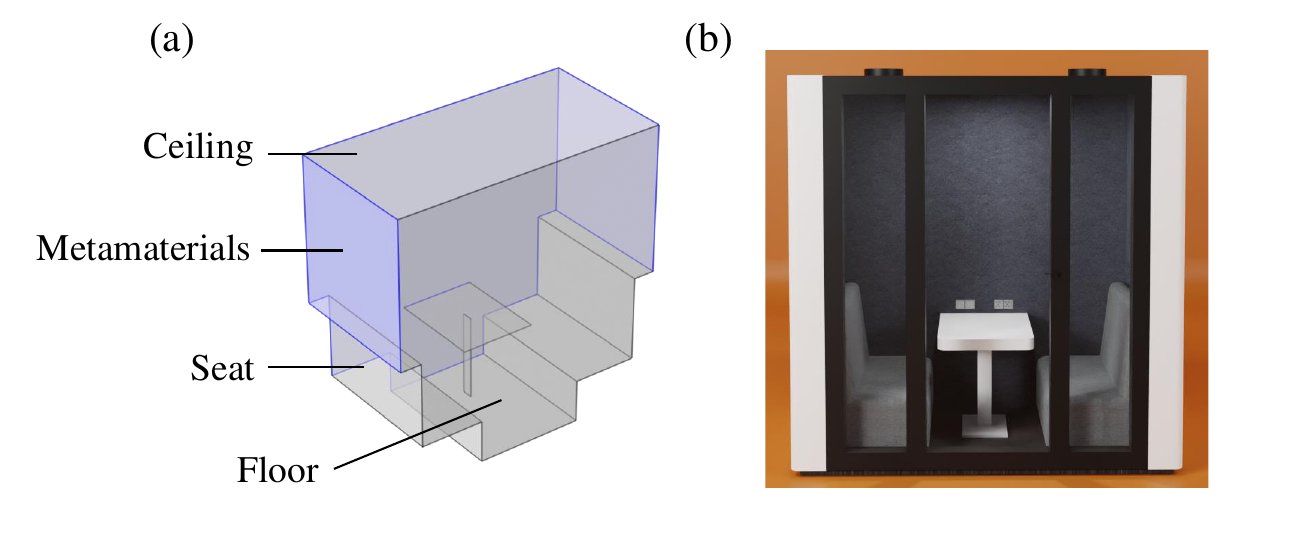}
	\caption{Small room model installed with acoustic metamaterials. (\textbf{a}) The material layout of the walls of the room. We assume the left, right and backward walls are covered with metamaterials. (\textbf{b}) A rendered schematic of the geometry of the proposed room.\label{fig1}}
\end{figure}

{
This study examines a small room with internal dimensions of approximately 1.1 $\times$ 2.1 $\times$ 2.1 m$^3$, which contains two seats and one table [Figure~\ref{fig1}(b)]. The room’s ceiling and floor are covered by polyester fibreboard, while the front wall with a door is made of transparent glass. The other three walls have covers that can be replaced with different materials, ranging from rigid walls and acoustic foam to metamaterial absorbers. The metamaterial is designed to have a specific impedance that matches the target reverberation time in the room.
}

\begin{figure}[b!]
	\includegraphics[width=13cm]{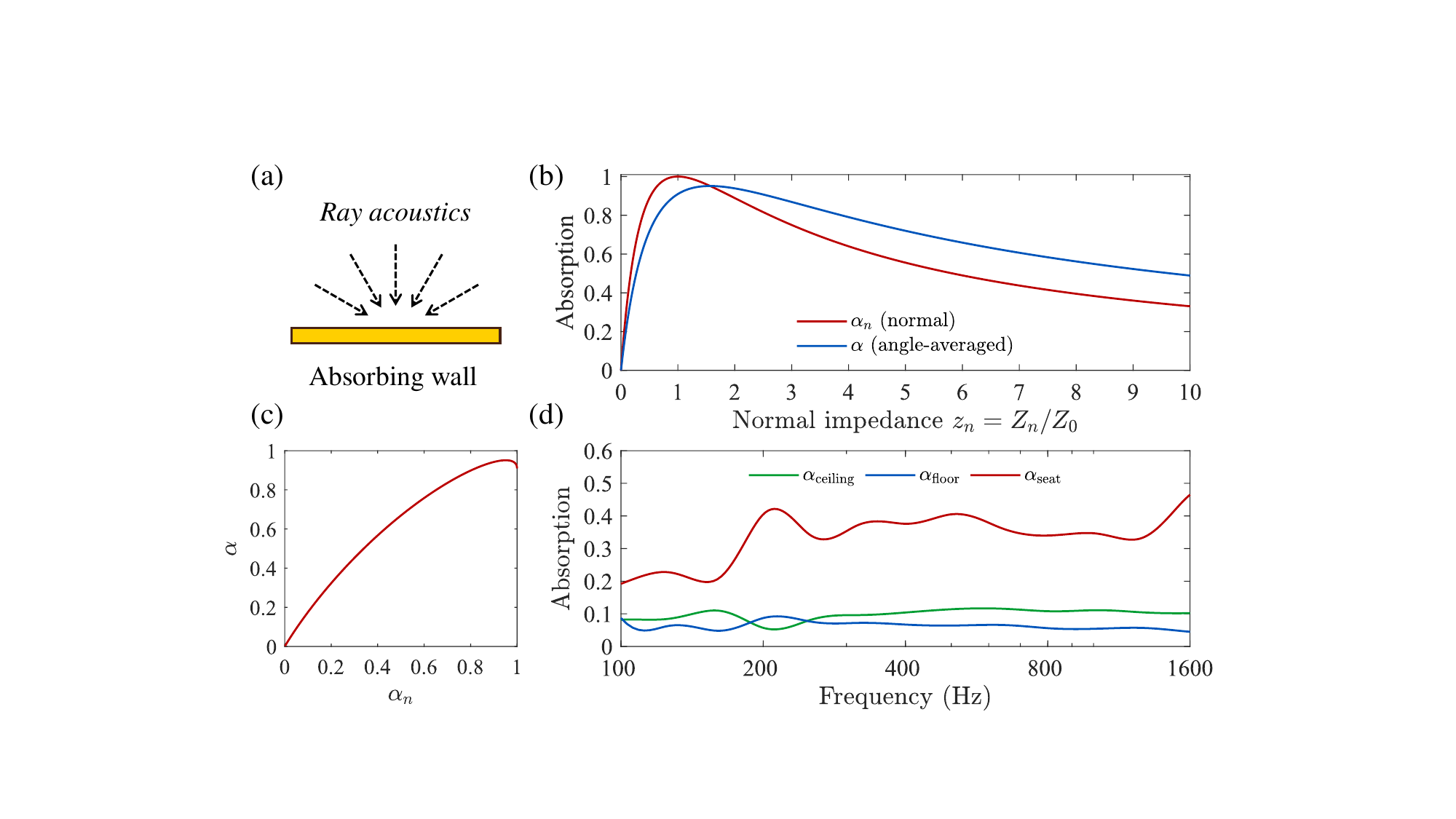}
	\centering
	\caption{Room walls with absorption. (\textbf{a}) The assumption of ray acoustics. The incident probability of sound rays from different directions is equal. (\textbf{b}) The comparison between the absorption coefficients of normal incidence ($\alpha_n$) and angle-averaged assumption ($\alpha$), which are plotted as functions of normalized surface impedance. (\textbf{c}) Conversion relation between $\alpha_n$ and $\alpha$. (\textbf{d}) The adopted angle-averaged absorption data of common ceiling, floor and seat \cite{aspock2020benchmark}. \label{fig2}}
\end{figure}

{
A normal impedance kernel, $Z_n(x,t;x',t')$, relates the local sound pressure on the surface, $p(x,t)$, to the past and present air molecule velocity, $v_n(x',t')$, along the normal direction at the same and nearby locations by $p(x,t)=\int Z_n(x,t;x',t')v_n(x',t')dx'dt'$. For metamaterials with resonators that only expose a \emph{subwavelength} face to sound, the velocity's contributions at the surrounding locations can be ignored. Thus, $Z_n$ becomes a complex function of frequency that can be expressed through a summation of Lorentz forms \cite{yang2017sound},
\begin{align*}
	\nonumber
	Z_n(\omega)&=\frac{i}{\omega}\left(\frac{\phi}{M}
	\sum_{m=1}^M\frac{\chi_m}{\omega_m^2-\omega^2-i\omega\beta}\right)^{-1}\\\nonumber
	&=\frac{i}{\omega}\left\{\frac{\phi}{M}
	\sum_{m=1}^M\left[\frac{\chi_m(\omega_m^2-\omega^2)}{(\omega_m^2-\omega^2)^2+\omega^2\beta^2}
	+\frac{i\chi_m\omega\beta}{(\omega_m^2-\omega^2)^2+\omega^2\beta^2}\right]\right\}^{-1},
\end{align*}
in which $\beta$ is the damping coefficient, $\omega$ is the angular frequency, $\chi_m$ and $\omega_m$ are the resonance strength and angular frequency of the $m$-th resonator, respectively, $M$ denotes the resonators' total number, and the surface porosity $\phi$ is the ratio of resonators' total opening area to the area exposed to the incident sound.  For simplicity and generality, here we ignore the higher-order modes of the resonators, which can be corrected by a more exact treatment as shown in Section ~\ref{sec:4} \cite{yang2017optimal}.  Broadband metamaterials usually have a high modal density per unit frequency.  Therefore, since the first term in the above summation changes sign from negative to positive around each resonant frequency, the summation of all the modes tends to cancel out, leading to a negligible net result.  In contrast, since the second term of the summation is always positive, the contributions of all modes are cumulatively added to each other, resulting in a relatively large value.  Thus, unlike traditional porous materials such as acoustic foam, the impedance of a broadband acoustic metamaterial is almost purely real [as illustrated in the inset of Fig.~\ref{fig5}(c)] and can be approximated in terms of a real-valued integral,
\begin{equation}
	Z_n(\omega)\simeq\frac{1}{\omega}\left[\frac{\phi}{M}\int_{\omega_1}^{\omega_M}
	\frac{\chi(x)\omega\beta}{(x^2-\omega^2)^2+\omega^2\beta^2}\mathcal{M}_d(x)dx\right]^{-1},
\end{equation}
here $\mathcal{M}_d(\omega)\equiv dm/d\omega$ is the modal density, and $\omega_{1(M)}$ is the lowest (highest) frequency of the resonances in considerations.  One can, therefore, customize $Z(\omega)$ by designing the values of $\chi(\omega)\mathcal{M}_d(\omega)$.

In the following, the absorption (or impedance) data of the ceiling, floor, and seat were obtained from Ref. \cite{aspock2020benchmark} [see Figure \ref{fig2}(d)] and the surface of table and glass were treated as rigid. The classical Eyring equation can use these parameters to estimate the real-valued impedance for metamaterial design based on a desired reverberation time when the frequency is above the reverberation limit.  To model the low-frequency range, where the Eyring equation is not applicable, we will employ the direct finite element method (FEM) simulation in COMSOL Multiphysics software to establish the relationship between the target metamaterial impedance and the desired volume-averaged sound decay rate (instead of the classically defined reverberation time).
}

\section{Theoretical Framework: Reverberation Control \label{sec:3}}

\subsection{Transition from resonance to reverberation}

{
The resonance modes of a room are typically sparse at low frequencies and denser at higher frequencies. When the modal density is sufficiently high, the contributions from different resonance modes will overlap significantly, resulting in a spatially uniform and fully reverberant sound field. The transition from the resonant to the reverberant condition is not clear-cut. Therefore, we use FEM to directly simulate this transition in the sample room.
}

\begin{figure}[b!]
	\includegraphics[width=10cm]{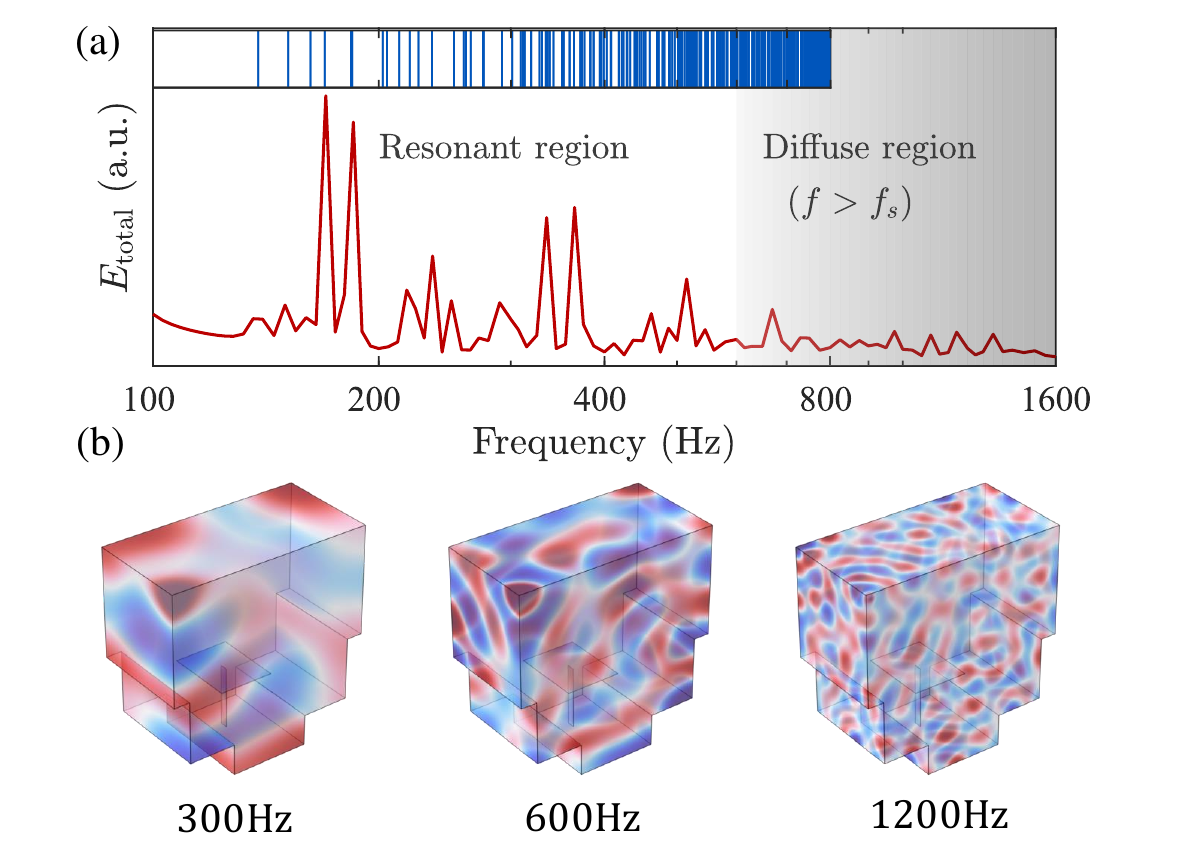}
	\centering
	\caption{Division of room characteristics: resonant region or diffuse region. (\textbf{a}) The total energy of the room as a function of frequency. The gradient gray color represents the transition from resonant to diffusive region. The inset figure displays the distribution of the eigenfrequencies below $f_s$ (real parts). (\textbf{b}) The pressure modal profiles under $ 300\,\mathrm{Hz} $, $ 600\,\mathrm{Hz} $ and $ 1200\,\mathrm{Hz} $. All the results above are without the installation of metamaterial absorbing boundaries on the walls. \label{fig3}}
\end{figure}  

In the simulation, we { used a sound monopole source on the top of the room (near the ceiling) to excite more diverse modes and} calculated the room's energy response curves at different frequencies [Figure \ref{fig3}(a)].  The total energy $E_\mathrm{total}$ was estimated based on the volume integral of sound energy density \cite{qu2020minimizing}.  { As shown in Figure \ref{fig3}(a), for an untreated room with rigid walls on three sides, $E_\text{total}$ fluctuates at low frequencies and becomes relatively smooth when the frequency is high.  This transition occurs around the Schroeder frequency, which is defined by $f_s=\alpha\sqrt{T_{60}(f_s)/V}$ with $V$ being the room's volume, $T_\text{60}$ the reverberation time, and $\alpha$ ranging from 2000[m/s]$^{\frac{3}{2}}$ to 4000[m/s]$^{\frac{3}{2}}$ \cite{noxon1987controlled}. In the current case, we took the commonly accepted value ($\alpha=2000\mathrm{[m/s]}^{\frac{3}{2}}$), thus $f_s=600$ Hz. The transition can also be observed from the eigenfrequencies shown in the inset of Figure 3(a) by blue vertical lines. The region of high modal density (grey) coincides with the smooth response frequency range, which is above $f_s$.}  Figure \ref{fig3}(b) displays the sound pressure field patterns at $300\,\mathrm{Hz}$, $600\,\mathrm{Hz}$ and $1200\,\mathrm{Hz}$, {which illustrate the transition of the overall field to a more reverberant and uniform pattern}.

{ The computational resources required for numerical simulation increase rapidly as the frequency increases. Therefore, we will only perform full-wave simulation for the resonance and transition regions and use conventional ray acoustics theory to calculate higher frequencies.}

\subsection{Ray Acoustics Assumption and Reverberation Time}

In room acoustics, ray acoustics assumes that when sound waves hit the surface of a wall, the probabilities of incidence at different angles are equal [Figure \ref{fig2}(a)]. Provided that the normal impedance of a wall is $Z_n$, according to the principle of the interference of incident and reflected plane waves, we know that the absorption coefficient for normal incidence is $\alpha_n=1-|(z_n-1)/(z_n+1)|^2$, where $z_n = Z_n(f)/Z_0$ and $Z_0$ is air's characteristic impedance. {  For broadband acoustic metamaterial with an almost real-valued impedance, if we further take the impedance change under different incident angles into consideration, effectively, there can be an angle-averaged version of the absorption \cite{kuttruff2012room,qu2020minimizing}
\begin{equation}\label{eq:angle_alpha}
	\alpha(f) = \frac{8}{{z_n}^2}\left(1+ z_n - \frac{1}{1+ z_n} - 2\ln(1+ z_n)\right).
\end{equation}
A comparison of $\alpha_n(f)$ and $\alpha(f)$ is plotted in Figure \ref{fig2}(b).} The angle-averaged absorption coefficient can never reach $ 100\% $, and there are differences between the two definitions of absorption coefficient in terms of optimal impedance: for normal incidence, impedance matching is always desired case, while for maximum angle-averaged absorption, the required impedance is slightly greater than that of air \cite{qu2020minimizing}. We plotted the two absorption coefficient definitions in Figure \ref{fig2}(c) to avoid confusion and facilitate the following conversion.

By definition, after the sources stop their radiation, the reverberation time $T_\mathrm{60}$ is the time required for the sound energy to decay by $ 10^{-6} $, approximately following the exponential relation
\begin{equation}\label{eq:decay}
	E_\mathrm{total}  = E_0 \exp\left(-\frac{\ln(10^6)}{T_\mathrm{60}}t\right).
\end{equation}
Based on ray acoustics, the most commonly used formula for $T_\mathrm{60}$ prediction is the  Eyring equation \cite{eyring1930reverberation,xiang2020generalization}:
\begin{equation}\label{eq:eyring}
	T_\mathrm{60} = \frac{0.163[\mathrm{s/m}]V}{-\sum_{i}S_i \ln(1-\alpha_i)},
\end{equation}
where $S_i$ can be treated as the surface area of $i^\mathrm{th}$ wall (e.g., $i$ can be ceiling, floor, seat or metamaterial). $\alpha_i$ is the $i^\mathrm{th}$ angle-averaged absorption coefficient.  It should be noted that Norris, Eyring, Schuster and Waitzmann all independently derived Equation (\ref{eq:eyring}). Although there are many more complex versions of $ T_\mathrm{60} $ prediction theory (such as when considering the coupling of two or more rooms \cite{eyring1931reverberation,schroeder1965new,billon2006use}), Eyring equation is still the most simple but effective description that is also applicable for higher absorption \cite{prawda2022calibrating}.

To control the reverberation time within a reasonable range, we set the target reverberation time to $\bar{T}_\mathrm{60} = 0.1\,\mathrm{s} $, which needs to be maintained as constant as possible over a wide frequency range. This time is the recommended value for high speech clarity in recording studios \cite{parker2009good} {and can be modified to other values due to the flexibility of the reverse-design approach.}  In general, the optimal reverberation time is linearly proportional to the logarithm of the room volume \cite{LONG2014829} $\bar{T}_\mathrm{60} \propto \ln(V)$. {Hence, optimal $T_\mathrm{60}$ is relatively small for small rooms, thus requiring higher absorption than that of large rooms \cite{mei2012experimental,d2020acoustic}.} 

By plugging this target value into the Eyring equation, we can obtain the following relationship,
\begin{equation}
	-\sum_{i} S_i \ln(1-\alpha_i)= \frac{0.163[\mathrm{s/m}]V}{\bar{T}_\mathrm{60}},
\end{equation}
yielding the target absorption for metamaterials
\begin{equation} \label{eq:t_ey}
	\alpha_\mathrm{meta}(f) = 1-\exp\left[ - \frac{0.163[\mathrm{s/m}]V}{ S_\mathrm{meta} \bar{T}_\mathrm{60}} -\sum_{i\neq \mathrm{meta}} \frac{S_i}{S_\mathrm{meta}} \ln(1-\alpha_i(f))\right] .
\end{equation}
For the ease of designing normal impedance, we should convert $\alpha_\mathrm{meta}(f) \to \alpha^\mathrm{meta}_{n}(f)$ according to the relation displayed in Figure \ref{fig2}(c). 

\begin{figure}[b!]
	\centering
	\includegraphics[width=14cm]{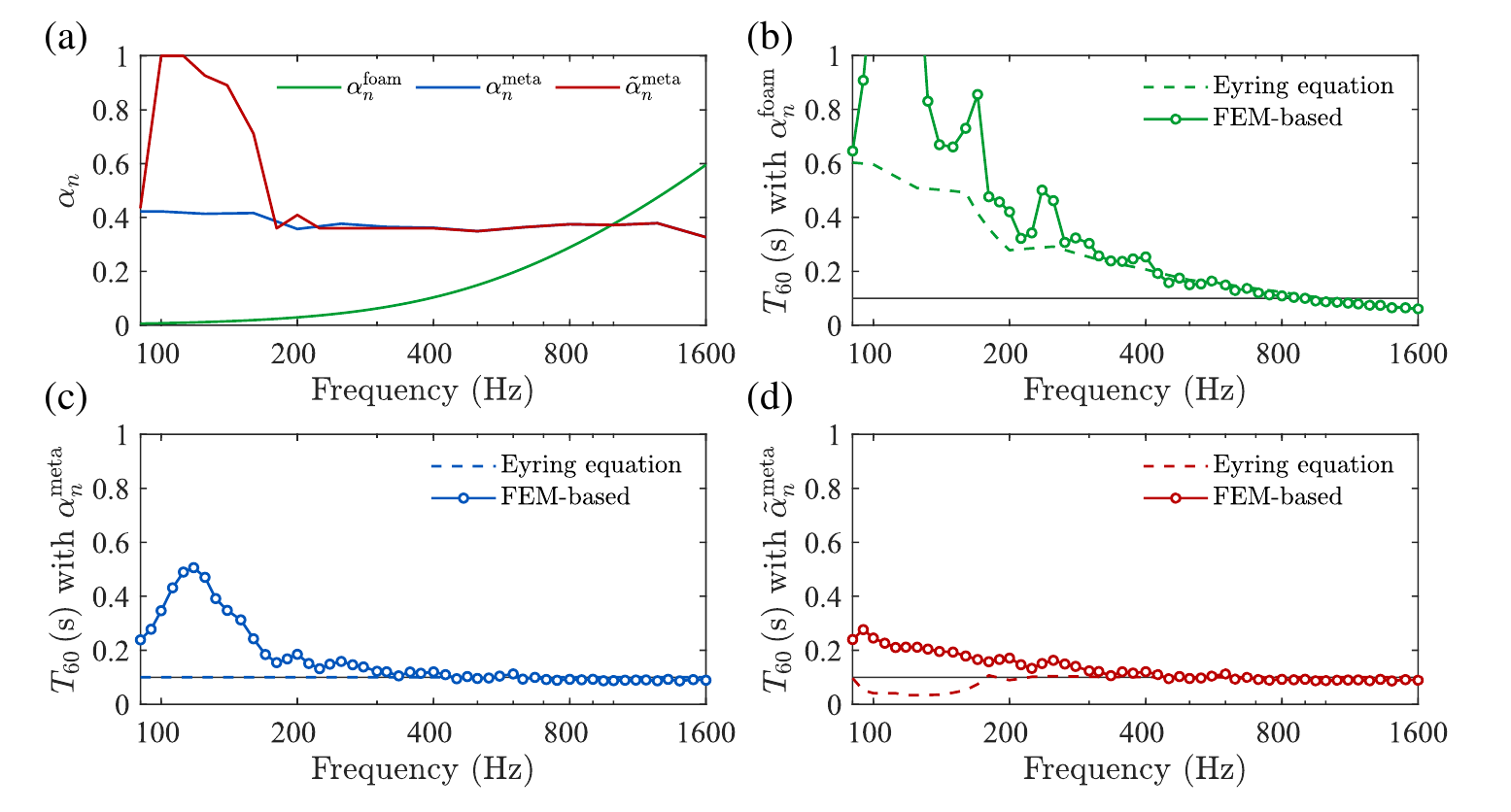}
	\caption{Simulation results for reverberation control. (\textbf{a}) Absorption target $\alpha_{n}^\mathrm{meta}$ (based on ray acoustic theory) and $\tilde{\alpha}_{n}^\mathrm{meta}$ corrected by FEM. Also, absorption of foam $\alpha_{n}^\mathrm{foam}$ based on the Johnson-Champoux-Allard (JCA) model is also presented. (\textbf{b-d}) $T_\mathrm{60}$ spectra with $\alpha_{n}^\mathrm{foam}$ (green), $\alpha_{n}^\mathrm{meta}$ (blue) and $\tilde{\alpha}_{n}^\mathrm{meta}$ (red). Circles are the results from the FEM simulation, while the dashed lines are from the ray-acoustics-based Eyring equation. {The difference between dashed lines and circles is caused by the failure of Eyring equation in the low-frequency region. Hence, modal reverberation time should be calculated by the FEM-based method.}\label{fig4}}
\end{figure}

{
Despite the limited validity of Equation (\ref{eq:t_ey}), we still plot its results over the entire frequency band in the blue line in Figure \ref{fig4}(a), using the parameters in Figure \ref{fig2}(d). The discrepancy between it and the FEM calculation results introduced later reflects the failure of ray acoustics.
}

\subsection{Sound Fields' Lifetime in Resonance Conditions}

{
At low frequencies, the room’s response is dominated by resonant behavior, which leads to a non-uniform sound field's lifetime across space. Therefore, the classical definition of $T_{60}$ is not applicable \cite{prato2016reverberation}. We propose a simple correction of the reverberation time by calculating the decay time of the averaged field instead of a uniform field decay time, which extends the definition of $T_{60}$ to lower frequency bands. In practice, the acoustic field averaging is based on FEM simulation results, and the $T_{60}$ results obtained should be consistent with what is from classical ray acoustics at high frequencies.
}

If we treat the room as a resonant cavity, its quality factor \cite{jackson2021classical} is defined by the ratio of the energy stored in the resonator to the energy dissipated per time cycle
\begin{equation}\label{eq:Q}
	Q = -\omega\frac{E_\mathrm{total}}{\partial_t E_\mathrm{total}},
\end{equation}
where{
\begin{align}
	&E_\mathrm{total} = \int_{V} \frac{1}{4}\left( \rho_0 |\mathbf{v}|^2 +\frac{1}{\kappa_0} |p|^2\right) d V,\label{eq:Q1}\\
	&\partial_t E_\mathrm{total} = \int_{S} \left( \frac{1}{2} p^* \mathbf{v}\right)  \cdot\mathbf{n} d S. \label{eq:Q2}
\end{align}}
Here, $\rho_0,\,\kappa_0$ are the air density and bulk modulus. $V$ is the room's interior space, while $S$ is the surface of the walls. The factors [$1/4$ in Equation (\ref{eq:Q1}) and $1/2$ in Equation (\ref{eq:Q2})] come from the time average of the complex fields, i.e., steady states are considered here. We consider frequency-domain solutions to reduce computation complexity as an alternative to time-domain simulations. Acoustic pressure field $ p $ and particle velocity $ \mathbf{v} $ as a function of three-dimensional space can be extracted from FEM simulation.

From Equation (\ref{eq:Q}), we can solve the differential equation and obtain $E_\mathrm{total}  =E_0 \exp\left(-\frac{\omega}{Q}t\right)  $. By comparing it with Equation (\ref{eq:decay}), we can solve $E_\mathrm{total}/E_0 = 10^{-6} $ and obtain $T_{60} = Q \ln(10^6)/\omega$. By inserting the definition in Equation (\ref{eq:Q}) into the above expression, we have
\begin{equation}\label{eq:T60_FEM}
	T_{60} = -\frac{E_\mathrm{total}}{\partial_t E_\mathrm{total}} \ln(10^6).
\end{equation}
Therefore, by FEM, we can visualize the impact of metamaterials by simulating with different frequencies and absorption spectra to obtain reverberation time as a guiding map [Figure \ref{fig5}(a)]. By searching for the absorption coefficient closest to the target $\bar{T}_{60} = 0.1\,\mathrm{s}$ for each frequency, we can obtain the optimal absorption spectrum $\tilde{\alpha}_{n}^\mathrm{meta}$ based on FEM [see red line in Figure \ref{fig4}(a) and dashed line in Figure \ref{fig5}(a)]. {It can be seen,} $\alpha_{n}^\mathrm{meta}$ and $\tilde{\alpha}_{n}^\mathrm{meta}$ are quite different below $200\,\mathrm{Hz}$ but beyond this frequency, the two almost coincide. Therefore, FEM provides a corrected optimal absorption spectrum, which we will use as the target for designing the metamaterial. In fact, $\tilde{\alpha}_{n}^\mathrm{meta}$ is dependent on the initial setting of the sound source, which can determine the relative strengths of different modes. It should be noted that the subsequent designs we present here are custom-made for the current setting of the room.

In Figure \ref{fig4}(c), we have shown the results of incorporating the absorption design based on the Eyring formula into the FEM model: at high frequencies, this design is basically accurate, and $ {T}_{60} $ is very close to the target value of $ 0.1\,\mathrm{s} $. In the low-frequency region below $ 200\,\mathrm{Hz} $, $ {T}_{60} $ is much higher than the target value due to the reduced absorption of other materials in the room and the gradually emerging resonant characteristics of the room. Stronger absorption is needed to counteract this effect. In Figure \ref{fig4}(d), the results using the revised target spectrum $\tilde{\alpha}_{n}^\mathrm{meta}$ are shown, and it can be seen that compared to Figure \ref{fig4}(c), the low-frequency part of the reverberation time is closer to the ideal value. { In summary, the difference between the dashed lines and circles in Figure \ref{fig4}(b-d) confirms the necessity of introducing FEM-based approach to evaluation modal reverberation time.}

We also examined traditional materials (absorbing foam) as a comparison [see green data in Figure \ref{fig4}(a-b)]. {Based on the Johnson-Champoux-Allard (JCA) model \cite{yang2017sound,johnson1987theory,champoux1991dynamic}, we set the equivalent parameters of the foam according to Ref. \cite{yang2017sound}, which includes the porosity ($\phi = 0.94$), flow resistivity ($R_\mathrm{f} = 32000 \,\mathrm{Pa\cdot s/m^{2}}$), tortuosity ($\alpha_\infty = 1.06$), viscous characteristic length ($\Lambda = 56\,\mathrm{\mu m}$) and thermal characteristic length ($\Lambda_\tau = 110\,\mathrm{\mu m}$).} We adjusted the foam thickness to $1.9\,\mathrm{cm}$ so that its absorption coefficient at $1\,\mathrm{kHz}$ would result in a reverberation time of $ 0.1\,\mathrm{s}$. However, after applying it to the FEM boundary conditions, we observe that the reverberation time deviated significantly from the target value below $1\,\mathrm{kHz}$, which highlights the significance of the proposed target absorption spectrum ($\tilde{\alpha}_{n}^\mathrm{meta}$). Beyond $1\,\mathrm{kHz}$, the reverberation time with foam should be lower than $ 0.1\,\mathrm{s} $. Therefore, a broadband and constant $ {T}_{60} $ spectrum cannot be achieved with traditional materials without a customized spectrum.

\section{Realization of Customized Metamaterial Absorption \label{sec:4}}

\begin{figure}[t!]
	\centering
	\includegraphics[width=13.6cm]{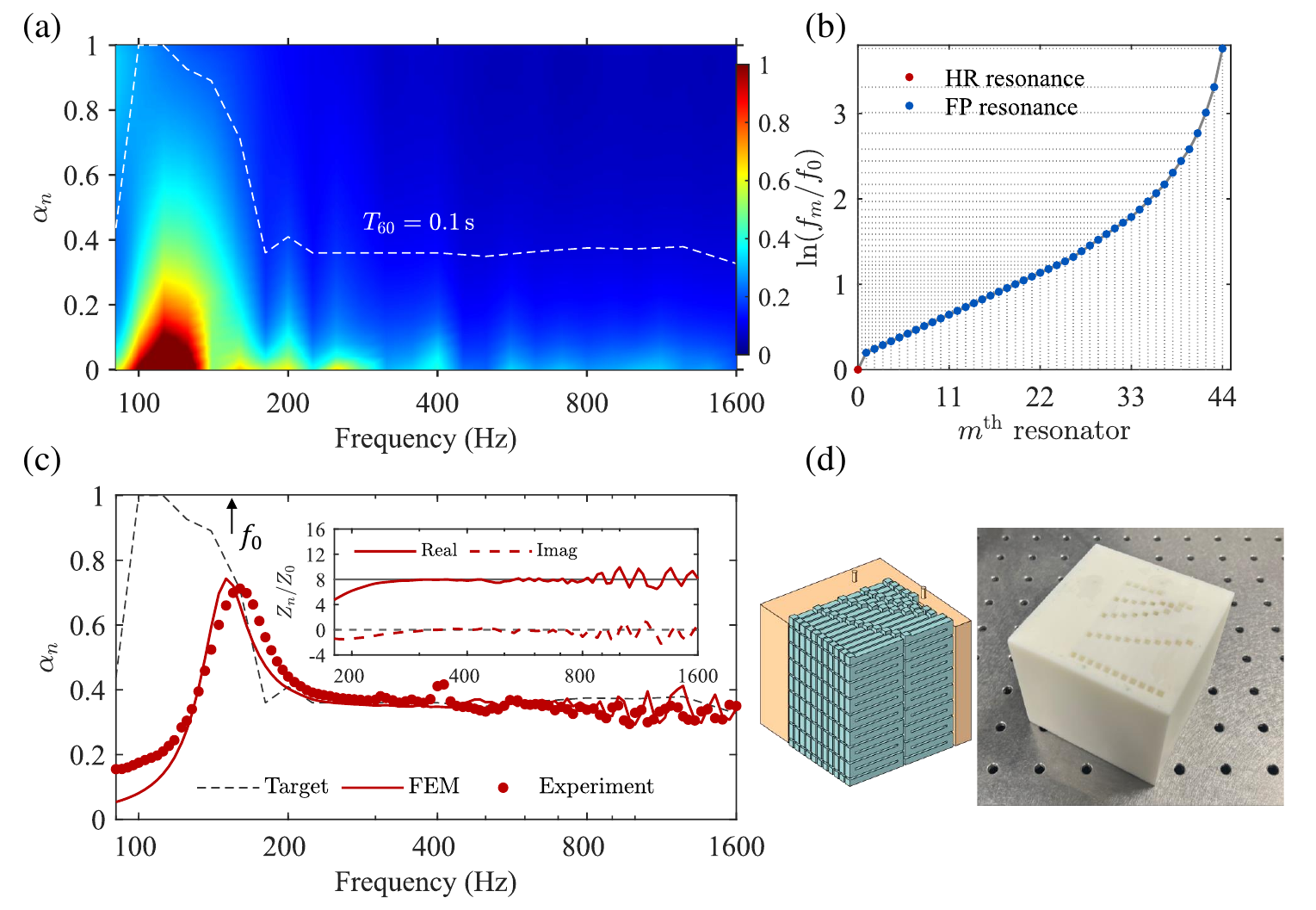}
	\caption{(\textbf{a}) Reverberation time as function of absorption and frequency. Dashed line shows the absorption curve that lets the resulting $ T_\mathrm{60} $ is closest to the target $\bar{T}_\mathrm{60} = 0.1\,\mathrm{s}$.  (\textbf{b}) The distribution of resonances $\{f_0,f_1,\cdots,f_N\}$ for target absorption ($ N=44 $). (\textbf{c}) FEM-predicted and measured absorption coefficients.  { The inset is the relevant impedance from numerical simulations.} (\textbf{d}) The corresponding sample schematic (left) and photo (right). \label{fig5}}
\end{figure}

Based on the previous discussion, the required absorption spectrum consists of a low-frequency absorption peak followed by a subsequent flat curve (with an absorption coefficient of approximately 0.4). Figure \ref{fig5}(c) shows this spectrum as a dashed line. According to previous work \cite{maa1998potential}, a suitable approach to design a low-frequency absorption peak is to use Helmholtz (HR) resonator, which can achieve a relatively pure single resonant mode. By adjusting the size and number of surface pores, as well as the volume of the rear cavity, impedance-matching conditions can be realized. The effect of the high-order modes is very weak, and it is in the high-frequency region (e.g., the second mode is already around 1550 Hz). So, the impedance of the HR resonator can be expressed in the Lorentzian function
\begin{equation}\label{eq:HR}
	Z_\mathrm{HR} \cong \frac{1}{-i\omega} \left( \frac{\chi_0}{\omega_0^2-\omega^2-i\omega\beta_0} \right) ^{-1},
\end{equation}
where the coefficients $\chi_0,\,\beta_0$ can be obtained by comparing Equation (\ref{eq:HR}) with the impedance function given by transfer matrix method \cite{maa1998potential}. The resonance frequency can be approximately determined by $\omega_0 = 2\pi f_0 = c\sqrt{S_\mathrm{HR}/(V_\mathrm{cavity}\tau)} $, where the aperture area $ S_\mathrm{HR} = 2 \pi (D_\mathrm{HR}/2)^2$. We set two HR necks with the diameter $D_\mathrm{HR} = 2\,\mathrm{mm}$ and neck length $\tau = 5\,\mathrm{mm}$. The cavity volume $V_\mathrm{cavity}\cong 1.2\times10^{-4}\,\mathrm{m}^3$. In this way, $f_0\cong175\,\mathrm{Hz}$ [see the red dot in Figure \ref{fig5}(b) and the peak location in Figure \ref{fig5}(c)]. The designed HR resonator is plotted with orange color in Figure \ref{fig5}(d).

Due to the availability of rich higher-order resonances, integrated Fabry–Pérot (FP) resonators \cite{yang2017sound,yang2017optimal,yang2018integration} are suitable for the absorption design with a flat smooth feature, whose overall impedance is
\begin{equation}\label{eq:FP}
	Z_\mathrm{FP} = \frac{1}{-i\omega} \left(\sum_{m=1}^{N} \sum_{k=1}^{\infty} \frac{\chi_m}{(2k-1)^2\omega_m^2-\omega^2-i\omega\beta_m} \right) ^{-1},
\end{equation}
where the coefficient $\chi_m = 2/(\rho_0 l_m)$ ($l_m$ is the length of $m_\mathrm{th}$ FP resonator), and the dissipation factor $\beta_m$ can be determined by the ratio between viscous boundary layer thickness and FP tube cross-section size ($w = 3.2\,\mathrm{mm}$ here). The total number of FP resonators $N=44$. Detailed discussions on how to derive Equations (\ref{eq:HR}) and (\ref{eq:FP}) for impedance design can be found in Ref. \cite{qu2022broadband}. In this case, the final overall impedance $ Z_n $ can be extracted from
\begin{equation}
	\frac{S_\mathrm{tube}}{Z_n} = \frac{S_\mathrm{HR}}{Z_\mathrm{HR}} +\frac{S_\mathrm{FP}}{Z_\mathrm{FP}},
\end{equation}
where the cross-section area of the impedance tube $ S_\mathrm{tube} = 7\mathrm{cm}\,\times\, 7\mathrm{cm} $. Or, $ S_\mathrm{tube}$ can be regarded as the horizontal area of a periodic unit (under the normal incidence of a plane wave). The aperture area of FP resonator $ S_\mathrm{FP} = N w^2$. The FP resonance design follows the design scheme described in Ref. \cite{yang2017optimal,qu2020minimizing} to achieve $\alpha_{n}\cong 0.4$ [thus $Z_n \cong 8 Z_0$], which contributes to the blue dots in Figure \ref{fig5}(b). As shown in Figure \ref{fig5}(d), the FP tubes (blue part) are coiled up to form a compact structure (together with the HR resonator). The left schematic is the air part that supports resonant modes. A practical sample fabricated by 3D-printing technology is displayed on the right side. The overall sample has a thickness of $6.8\,\mathrm{cm}  $ and its absorption data [red dots in Figure \ref{sec:5}(c)] agrees well with the targeted dashed line when $f>f_0$. Below $f_0$, in principle, we can introduce more resonances to cover the low-frequency band. Still, since the available space is limited, so under the current condition, we only achieve $f_0 = 175 \,\mathrm{Hz}$. The thickness-absorption trade-off can be well captured by the causality constraint relation in diverse wave systems \cite{yang2017optimal,qu2021conceptual,qu2022underwater}.

\section{Experimental Verification in a Small Room \label{sec:5}}
\begin{figure}[t!]
	\includegraphics[width=13.6cm]{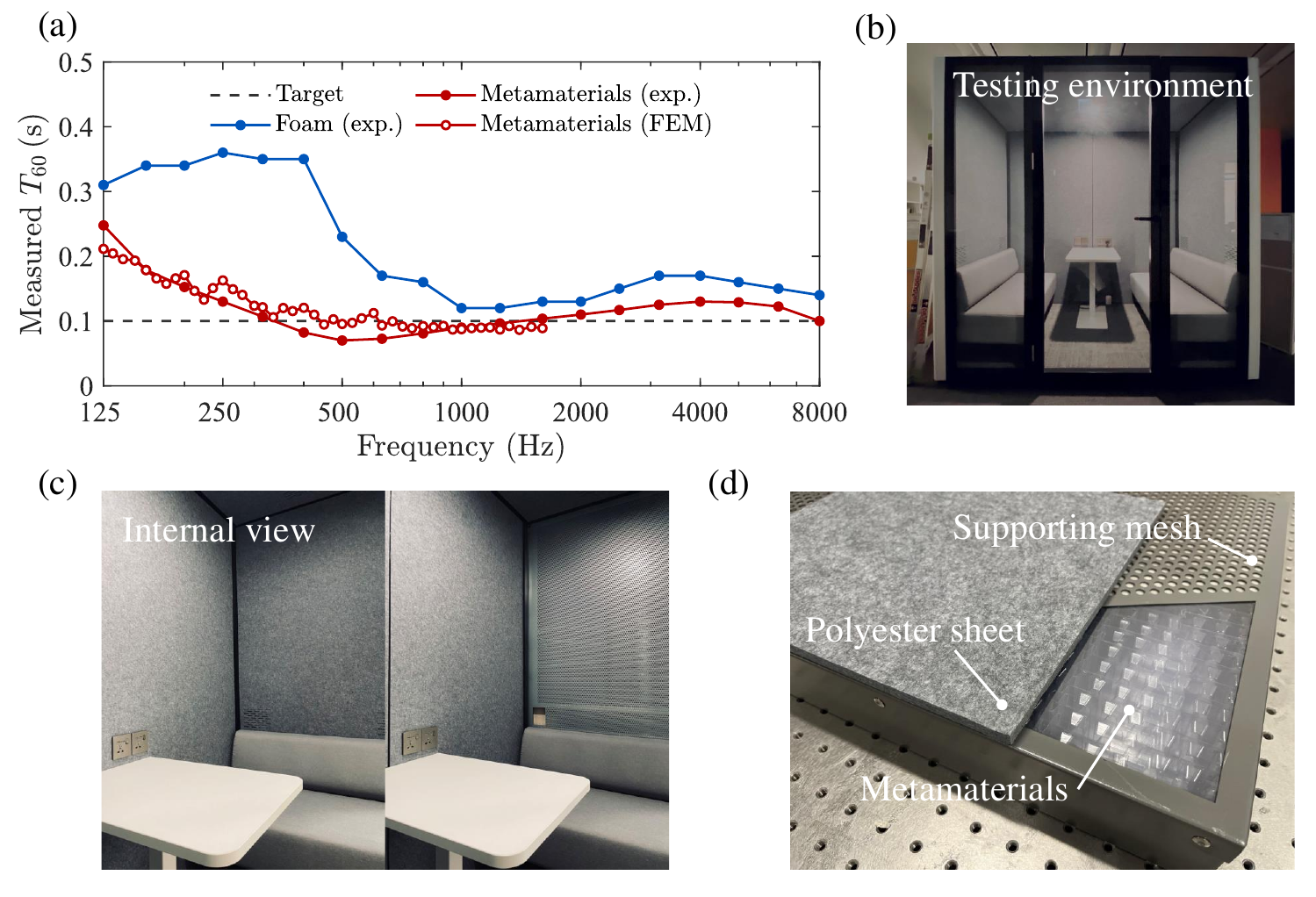}
	\caption{Reverberation time control performance in a small room. (\textbf{a}) Measured $T_\mathrm{60}$ spectra of the proposed metamaterial and a commercial foam. {The $T_\mathrm{60}$-evaluation by FEM-based method is also displayed for comparison [the same data in Figure \ref{fig4}(d)].} (\textbf{b}) A photo of the testing environment. The metamaterial panels on the walls [the same layout as Figure \ref{fig1}(a)] are covered by a polyester sheet. (\textbf{c}) The internal views of the small room with and without the polyester sheet covering one side of the walls. (\textbf{d}) The detailed view of the used acoustic panel with supporting mesh (without acoustic functions) and polyester sheet (to increase the dissipation). \label{fig6}}
\end{figure}

In the previous section, we demonstrated through simulation and experiment that metamaterials can achieve the desired absorption spectrum. Here, on the one hand, we further show that the customized metamaterials can function in a real room. The necessity of measuring the reverberation time in a room is that impedance tube testing can only characterize the effect of normal incidence and cannot include the interaction between metamaterials and the complex reverberating modes of the room. On the other hand, even if the metamaterials can mimic the target spectral shape, we still need to verify through room testing that the optimal absorption spectrum we derived is correct. Before we conducted the measurements, we modified the metamaterial by embedding it underneath a polyester fiber sheet as a dissipative material [see Figure \ref{fig6}(c-d)] for smoothing high-frequency oscillations of the absorption coefficient \cite{yang2017sound,yang2017optimal}.  Additionally, we utilized a molding process suitable for large-scale manufacturing as the sample fabrication technique.

The measurements were guided by ISO 3382 standards \cite{ISO}. We used NTi-XL2 acoustic analyzer from NTi Audio AG company, which supports direct reverberation time measurements. The signal receiver was placed in the center of the room [see Figure \ref{fig6}(b)], with a broadband pulse sound source placed at the top of the room to excite diverse modes and prevent the strong resonant response of any specific single mode. {The source power, which has the estimated power of $10^{-3}\,\mathrm{W}$, is high enough to maintain a high signal-to-noise ratio and the insertion loss between the room interior and outside space is about 20 dB. The built-in algorithm (Schroeder method) of the analyzer can automatically fit the pressure data received to output the reverberation time. This algorithm takes advantage of the ensemble average being equal to the backward integration of the squared impulse response, to obtain satisfactory accuracy especially for low decaying time. According to the ISO \cite{ISO}, a recommended quantitative method for checking the feasibility is to examine the following condition: $B_w \times T_\mathrm{60} >4$, where $B_w$ is the filter bandwidth, and $T_\mathrm{60}$ is the estimated reverberation time. In our case, $B_w \times T_\mathrm{60} \cong 7.25$ because at $125 \,\mathrm{Hz}$ (the lowest frequency in our experiment), $B_w$ for 1/3-octave band\footnote{For 1/3-octave band with center frequency $f_c$, the bandwidth $B_w$ is $(2^{\frac{1}{6}}-2^{-\frac{1}{6}}) f_c\cong 23\% f_c$.} is $29\,\mathrm{Hz}$ and $ T_\mathrm{60} \cong 0.25\,\mathrm{s}$.} As a comparison, we also tested a commercial foam\footnote{The foam is the interior material of an acoustic booth product XR-M F4 from Sound Box company.} using the same method. The measurement results in Figure \ref{fig6}(a) showed that for the target $T_\mathrm{60} = 0.1\,\mathrm{s}$ (dashed line), the foam experienced a significant increase in reverberation time at low frequencies (blue points). By contrast, the metamaterial exhibited a more uniform response, fluctuating near $ 0.1\,\mathrm{s} $, and this performance could be maintained from $ 250\,\mathrm{Hz} $ up to $ 8\,\mathrm{kHz} $ (red points). {In Figure \ref{fig6}(a), we also plot red circles for the FEM simulated results [from Figure \ref{fig4}(d)] as a comparison.  The highest simulation frequency is limited to 1600 Hz due to the limited computational resources. The good agreement between the experimental and simulated results within this frequency band evidences the effectiveness of the proposed reverse design method of metamaterials in small rooms.}

\section{Concluding Remarks}
\label{sec:6}

In summary, designing the acoustic environment of small rooms is difficult due to a significant portion of the audible frequency band being outside the realm of traditional ray acoustics. This results in high non-uniformity in both frequency and space, making it necessary to implement advanced absorption technologies that are highly efficient and customizable across the frequency spectrum in order to achieve a uniform and comfortable acoustic environment. Another challenge is extending the concept of reverberation time to cavity resonant region. In this article, we evaluate the lifetime of the acoustic signal in the resonant cavity by using its quality factor and combining it with the classical Eyring equation to calculate the reverberation time. We then reverse-engineer the required wall absorption spectrum for a target uniform reverberation time spectrum. The proposed approach involves utilizing customized broadband metamaterial absorption technology through an array of different resonators, which has shown promising results in both numerical simulations and experiments. This strategy of designing absorption materials in a reverse manner from the final acoustic requirements is scalable and can flexibly provide optimized acoustic environments for different types of rooms and purposes \cite{barron2009auditorium}, including small recording studios, listening rooms, phone booths, and even car cabins with high acoustic requirements in enclosed spaces. Furthermore, this design approach can be extended to larger spaces such as concert halls, lecture halls, and theaters, as an alternative to traditional acoustic absorption materials. Finally, it should be noted that the design in this paper was based on passive resonators, which has the advantages of stability and low cost. In the future, the passive customized materials can be used as a supplementary component of the active solutions \cite{cecchi2017room} to control the indoor sound field, in order to reduce costs and alleviate the difficulty of control algorithms at high frequencies.

\section*{CRediT authorship contribution statement}
\textbf{Sichao Qu:} Writing – original draft, Visualization, Methodology,
Investigation, Formal analysis, Software. \textbf{Min Yang:} Writing – review \& editing, Formal analysis, Supervision, Funding acquisition, Conceptualization. \textbf{Yunfei Xu:} Validation, Data curation, Investigation. \textbf{Songwen Xiao:} Project administration, Resources. \textbf{Nicholas X. Fang:} Writing - Review \& editing, Supervision, Validation.
\section*{Declaration of competing interest}
The authors declare no conflict of interest.
\section*{Data availability}
The data that support the findings of this study are available upon request from the corresponding author.



\bibliographystyle{elsarticle-num-names} 

\begin{thebibliography}{51}
	\expandafter\ifx\csname natexlab\endcsname\relax\def\natexlab#1{#1}\fi
	\providecommand{\url}[1]{\texttt{#1}}
	\providecommand{\href}[2]{#2}
	\providecommand{\path}[1]{#1}
	\providecommand{\DOIprefix}{doi:}
	\providecommand{\ArXivprefix}{arXiv:}
	\providecommand{\URLprefix}{URL: }
	\providecommand{\Pubmedprefix}{pmid:}
	\providecommand{\doi}[1]{\href{http://dx.doi.org/#1}{\path{#1}}}
	\providecommand{\Pubmed}[1]{\href{pmid:#1}{\path{#1}}}
	\providecommand{\bibinfo}[2]{#2}
	\ifx\xfnm\relax \def\xfnm[#1]{\unskip,\space#1}\fi
	\bibitem[{Kuttruff and Mommertz(2012)}]{kuttruff2012room}
	\bibinfo{author}{H.~Kuttruff}, \bibinfo{author}{E.~Mommertz},
	\newblock \bibinfo{title}{Room acoustics},
	\newblock in: \bibinfo{booktitle}{Handbook of engineering acoustics},
	\bibinfo{publisher}{Springer}, \bibinfo{year}{2012}, pp.
	\bibinfo{pages}{239--267}.
	\bibitem[{Parker(2009)}]{parker2009good}
	\bibinfo{author}{B.~Parker}, \bibinfo{title}{Good vibrations: The physics of
		music}, \bibinfo{publisher}{JHU Press}, \bibinfo{year}{2009}.
	\bibitem[{Garai et~al.(2016)Garai, De~Cesaris, Morandi, and
		D’Orazio}]{garai2016sound}
	\bibinfo{author}{M.~Garai}, \bibinfo{author}{S.~De~Cesaris},
	\bibinfo{author}{F.~Morandi}, \bibinfo{author}{D.~D’Orazio},
	\newblock \bibinfo{title}{Sound energy distribution in italian opera houses},
	\newblock in: \bibinfo{booktitle}{Proceedings of Meetings on Acoustics 22ICA},
	volume~\bibinfo{volume}{28}, \bibinfo{organization}{Acoustical Society of
		America}, \bibinfo{year}{2016}, p. \bibinfo{pages}{015019}.
	\bibitem[{Nolan et~al.(2020)Nolan, Berzborn, and
		Fernandez-Grande}]{nolan2020isotropy}
	\bibinfo{author}{M.~Nolan}, \bibinfo{author}{M.~Berzborn},
	\bibinfo{author}{E.~Fernandez-Grande},
	\newblock \bibinfo{title}{Isotropy in decaying reverberant sound fields},
	\newblock \bibinfo{journal}{The Journal of the Acoustical Society of America}
	\bibinfo{volume}{148} (\bibinfo{year}{2020}) \bibinfo{pages}{1077--1088}.
	\bibitem[{Sagartzazu et~al.(2008)Sagartzazu, Hervella-Nieto, and
		Pagalday}]{sagartzazu2008review}
	\bibinfo{author}{X.~Sagartzazu}, \bibinfo{author}{L.~Hervella-Nieto},
	\bibinfo{author}{J.~Pagalday},
	\newblock \bibinfo{title}{Review in sound absorbing materials},
	\newblock \bibinfo{journal}{Archives of Computational Methods in Engineering}
	\bibinfo{volume}{15} (\bibinfo{year}{2008}) \bibinfo{pages}{311--342}.
	\bibitem[{Shin and Kim(2020)}]{shin2020sound}
	\bibinfo{author}{H.-k. Shin}, \bibinfo{author}{K.-w. Kim},
	\newblock \bibinfo{title}{Sound absorbing ceiling to reduce heavy weight floor
		impact sound},
	\newblock \bibinfo{journal}{Building and Environment} \bibinfo{volume}{180}
	(\bibinfo{year}{2020}) \bibinfo{pages}{107058}.
	\bibitem[{Ma and Sheng(2016)}]{ma2016acoustic}
	\bibinfo{author}{G.~Ma}, \bibinfo{author}{P.~Sheng},
	\newblock \bibinfo{title}{Acoustic metamaterials: From local resonances to
		broad horizons},
	\newblock \bibinfo{journal}{Science advances} \bibinfo{volume}{2}
	(\bibinfo{year}{2016}) \bibinfo{pages}{e1501595}.
	\bibitem[{Yang and Sheng(2017)}]{yang2017sound}
	\bibinfo{author}{M.~Yang}, \bibinfo{author}{P.~Sheng},
	\newblock \bibinfo{title}{Sound absorption structures: From porous media to
		acoustic metamaterials},
	\newblock \bibinfo{journal}{Annual Review of Materials Research}
	\bibinfo{volume}{47} (\bibinfo{year}{2017}) \bibinfo{pages}{83--114}.
	\bibitem[{Cummer et~al.(2016)Cummer, Christensen, and
		Al{\`u}}]{cummer2016controlling}
	\bibinfo{author}{S.~A. Cummer}, \bibinfo{author}{J.~Christensen},
	\bibinfo{author}{A.~Al{\`u}},
	\newblock \bibinfo{title}{Controlling sound with acoustic metamaterials},
	\newblock \bibinfo{journal}{Nature Reviews Materials} \bibinfo{volume}{1}
	(\bibinfo{year}{2016}) \bibinfo{pages}{1--13}.
	\bibitem[{Gu et~al.(2021)Gu, Gao, Cao, Liu, Zhu, and Zhu}]{gu2021controlling}
	\bibinfo{author}{Z.~Gu}, \bibinfo{author}{H.~Gao}, \bibinfo{author}{P.-C. Cao},
	\bibinfo{author}{T.~Liu}, \bibinfo{author}{X.-F. Zhu},
	\bibinfo{author}{J.~Zhu},
	\newblock \bibinfo{title}{Controlling sound in non-hermitian acoustic systems},
	\newblock \bibinfo{journal}{Physical Review Applied} \bibinfo{volume}{16}
	(\bibinfo{year}{2021}) \bibinfo{pages}{057001}.
	\bibitem[{Xue et~al.(2022)Xue, Yang, and Zhang}]{xue2022topological}
	\bibinfo{author}{H.~Xue}, \bibinfo{author}{Y.~Yang},
	\bibinfo{author}{B.~Zhang},
	\newblock \bibinfo{title}{Topological acoustics},
	\newblock \bibinfo{journal}{Nature Reviews Materials}  (\bibinfo{year}{2022})
	\bibinfo{pages}{1--17}.
	\bibitem[{Jim{\'e}nez et~al.(2017)Jim{\'e}nez, Romero-Garc{\'\i}a, Pagneux, and
		Groby}]{jimenez2017rainbow}
	\bibinfo{author}{N.~Jim{\'e}nez}, \bibinfo{author}{V.~Romero-Garc{\'\i}a},
	\bibinfo{author}{V.~Pagneux}, \bibinfo{author}{J.-P. Groby},
	\newblock \bibinfo{title}{Rainbow-trapping absorbers: Broadband, perfect and
		asymmetric sound absorption by subwavelength panels for transmission
		problems},
	\newblock \bibinfo{journal}{Scientific reports} \bibinfo{volume}{7}
	(\bibinfo{year}{2017}) \bibinfo{pages}{13595}.
	\bibitem[{Jiang et~al.(2014)Jiang, Liang, Li, Zou, Yin, and
		Cheng}]{jiang2014ultra}
	\bibinfo{author}{X.~Jiang}, \bibinfo{author}{B.~Liang}, \bibinfo{author}{R.-q.
		Li}, \bibinfo{author}{X.-y. Zou}, \bibinfo{author}{L.-l. Yin},
	\bibinfo{author}{J.-c. Cheng},
	\newblock \bibinfo{title}{Ultra-broadband absorption by acoustic
		metamaterials},
	\newblock \bibinfo{journal}{Applied Physics Letters} \bibinfo{volume}{105}
	(\bibinfo{year}{2014}) \bibinfo{pages}{243505}.
	\bibitem[{Long et~al.(2019)Long, Shao, Liu, Cheng, and Liu}]{long2019broadband}
	\bibinfo{author}{H.~Long}, \bibinfo{author}{C.~Shao}, \bibinfo{author}{C.~Liu},
	\bibinfo{author}{Y.~Cheng}, \bibinfo{author}{X.~Liu},
	\newblock \bibinfo{title}{Broadband near-perfect absorption of low-frequency
		sound by subwavelength metasurface},
	\newblock \bibinfo{journal}{Applied Physics Letters} \bibinfo{volume}{115}
	(\bibinfo{year}{2019}) \bibinfo{pages}{103503}.
	\bibitem[{Yang et~al.(2017)Yang, Chen, Fu, and Sheng}]{yang2017optimal}
	\bibinfo{author}{M.~Yang}, \bibinfo{author}{S.~Chen}, \bibinfo{author}{C.~Fu},
	\bibinfo{author}{P.~Sheng},
	\newblock \bibinfo{title}{Optimal sound-absorbing structures},
	\newblock \bibinfo{journal}{Materials Horizons} \bibinfo{volume}{4}
	(\bibinfo{year}{2017}) \bibinfo{pages}{673--680}.
	\bibitem[{Rozanov(2000)}]{rozanov2000ultimate}
	\bibinfo{author}{K.~N. Rozanov},
	\newblock \bibinfo{title}{Ultimate thickness to bandwidth ratio of radar
		absorbers},
	\newblock \bibinfo{journal}{IEEE Transactions on Antennas and Propagation}
	\bibinfo{volume}{48} (\bibinfo{year}{2000}) \bibinfo{pages}{1230--1234}.
	\bibitem[{Acher et~al.(2009)Acher, Bernard, Mar{\'e}chal, Bardaine, and
		Levassort}]{acher2009fundamental}
	\bibinfo{author}{O.~Acher}, \bibinfo{author}{J.~Bernard},
	\bibinfo{author}{P.~Mar{\'e}chal}, \bibinfo{author}{A.~Bardaine},
	\bibinfo{author}{F.~Levassort},
	\newblock \bibinfo{title}{Fundamental constraints on the performance of
		broadband ultrasonic matching structures and absorbers},
	\newblock \bibinfo{journal}{The Journal of the Acoustical Society of America}
	\bibinfo{volume}{125} (\bibinfo{year}{2009}) \bibinfo{pages}{1995--2005}.
	\bibitem[{Meng et~al.(2022)Meng, Romero-Garc{\'\i}a, Gabard, Groby, Bricault,
		Goud{\'e}, and Sheng}]{meng2022fundamental}
	\bibinfo{author}{Y.~Meng}, \bibinfo{author}{V.~Romero-Garc{\'\i}a},
	\bibinfo{author}{G.~Gabard}, \bibinfo{author}{J.-P. Groby},
	\bibinfo{author}{C.~Bricault}, \bibinfo{author}{S.~Goud{\'e}},
	\bibinfo{author}{P.~Sheng},
	\newblock \bibinfo{title}{Fundamental constraints on broadband passive acoustic
		treatments in unidimensional scattering problems},
	\newblock \bibinfo{journal}{Proceedings of the Royal Society A}
	\bibinfo{volume}{478} (\bibinfo{year}{2022}) \bibinfo{pages}{20220287}.
	\bibitem[{Yang and Sheng(2018)}]{yang2018integration}
	\bibinfo{author}{M.~Yang}, \bibinfo{author}{P.~Sheng},
	\newblock \bibinfo{title}{An integration strategy for acoustic metamaterials to
		achieve absorption by design},
	\newblock \bibinfo{journal}{Applied Sciences} \bibinfo{volume}{8}
	(\bibinfo{year}{2018}) \bibinfo{pages}{1247}.
	\bibitem[{Qu and Sheng(2022)}]{qu2022microwave}
	\bibinfo{author}{S.~Qu}, \bibinfo{author}{P.~Sheng},
	\newblock \bibinfo{title}{Microwave and acoustic absorption metamaterials},
	\newblock \bibinfo{journal}{Physical Review Applied} \bibinfo{volume}{17}
	(\bibinfo{year}{2022}) \bibinfo{pages}{047001}.
	\bibitem[{Raman et~al.(2014)Raman, Anoma, Zhu, Rephaeli, and
		Fan}]{raman2014passive}
	\bibinfo{author}{A.~P. Raman}, \bibinfo{author}{M.~A. Anoma},
	\bibinfo{author}{L.~Zhu}, \bibinfo{author}{E.~Rephaeli},
	\bibinfo{author}{S.~Fan},
	\newblock \bibinfo{title}{Passive radiative cooling below ambient air
		temperature under direct sunlight},
	\newblock \bibinfo{journal}{Nature} \bibinfo{volume}{515}
	(\bibinfo{year}{2014}) \bibinfo{pages}{540--544}.
	\bibitem[{Li et~al.(2019)Li, Lin, Zhou, An, Li, Chi, Huang, Yang, Tso, Chao
		et~al.}]{li2019scalable}
	\bibinfo{author}{Y.~Li}, \bibinfo{author}{C.~Lin}, \bibinfo{author}{D.~Zhou},
	\bibinfo{author}{Y.~An}, \bibinfo{author}{D.~Li}, \bibinfo{author}{C.~Chi},
	\bibinfo{author}{H.~Huang}, \bibinfo{author}{S.~Yang}, \bibinfo{author}{C.~Y.
		Tso}, \bibinfo{author}{C.~Y. Chao}, et~al.,
	\newblock \bibinfo{title}{Scalable all-ceramic nanofilms as highly efficient
		and thermally stable selective solar absorbers},
	\newblock \bibinfo{journal}{Nano Energy} \bibinfo{volume}{64}
	(\bibinfo{year}{2019}) \bibinfo{pages}{103947}.
	\bibitem[{Wesemann et~al.(2019)Wesemann, Panchenko, Singh, Della~Gaspera,
		G{\'o}mez, Davis, and Roberts}]{wesemann2019selective}
	\bibinfo{author}{L.~Wesemann}, \bibinfo{author}{E.~Panchenko},
	\bibinfo{author}{K.~Singh}, \bibinfo{author}{E.~Della~Gaspera},
	\bibinfo{author}{D.~E. G{\'o}mez}, \bibinfo{author}{T.~J. Davis},
	\bibinfo{author}{A.~Roberts},
	\newblock \bibinfo{title}{Selective near-perfect absorbing mirror as a spatial
		frequency filter for optical image processing},
	\newblock \bibinfo{journal}{APL Photonics} \bibinfo{volume}{4}
	(\bibinfo{year}{2019}) \bibinfo{pages}{100801}.
	\bibitem[{Ma et~al.(2018)Ma, Fan, Sheng, and Fink}]{ma2018shaping}
	\bibinfo{author}{G.~Ma}, \bibinfo{author}{X.~Fan}, \bibinfo{author}{P.~Sheng},
	\bibinfo{author}{M.~Fink},
	\newblock \bibinfo{title}{Shaping reverberating sound fields with an actively
		tunable metasurface},
	\newblock \bibinfo{journal}{Proceedings of the National Academy of Sciences}
	\bibinfo{volume}{115} (\bibinfo{year}{2018}) \bibinfo{pages}{6638--6643}.
	\bibitem[{Qu and Sheng(2020)}]{qu2020minimizing}
	\bibinfo{author}{S.~Qu}, \bibinfo{author}{P.~Sheng},
	\newblock \bibinfo{title}{Minimizing indoor sound energy with tunable
		metamaterial surfaces},
	\newblock \bibinfo{journal}{Physical Review Applied} \bibinfo{volume}{14}
	(\bibinfo{year}{2020}) \bibinfo{pages}{034060}.
	\bibitem[{Wang et~al.(2022)Wang, del Hougne, and Ma}]{wang2022controlling}
	\bibinfo{author}{Q.~Wang}, \bibinfo{author}{P.~del Hougne},
	\bibinfo{author}{G.~Ma},
	\newblock \bibinfo{title}{Controlling the spatiotemporal response of transient
		reverberating sound},
	\newblock \bibinfo{journal}{Physical Review Applied} \bibinfo{volume}{17}
	(\bibinfo{year}{2022}) \bibinfo{pages}{044007}.
	\bibitem[{Kaplanis et~al.(2014)Kaplanis, Bech, Jensen, and van
		Waterschoot}]{kaplanis2014perception}
	\bibinfo{author}{N.~Kaplanis}, \bibinfo{author}{S.~Bech},
	\bibinfo{author}{S.~H. Jensen}, \bibinfo{author}{T.~van Waterschoot},
	\newblock \bibinfo{title}{Perception of reverberation in small rooms: a
		literature study},
	\newblock in: \bibinfo{booktitle}{Audio engineering society conference: 55th
		international conference: Spatial audio}, \bibinfo{organization}{Audio
		Engineering Society}, \bibinfo{year}{2014}.
	\bibitem[{Traer and McDermott(2016)}]{raer2016statistics}
	\bibinfo{author}{J.~Traer}, \bibinfo{author}{J.~H. McDermott},
	\newblock \bibinfo{title}{Statistics of natural reverberation enable perceptual
		separation of sound and space},
	\newblock \bibinfo{journal}{Proceedings of the National Academy of Sciences}
	\bibinfo{volume}{113} (\bibinfo{year}{2016}) \bibinfo{pages}{E7856--E7865}.
	\bibitem[{Schroeder(1987)}]{schroeder1987statistical}
	\bibinfo{author}{M.~R. Schroeder},
	\newblock \bibinfo{title}{Statistical parameters of the frequency response
		curves of large rooms},
	\newblock \bibinfo{journal}{Journal of the Audio Engineering Society}
	\bibinfo{volume}{35} (\bibinfo{year}{1987}) \bibinfo{pages}{299--306}.
	\bibitem[{Asp{\"o}ck et~al.(2020)Asp{\"o}ck, Vorl{\"a}nder, Brinkmann,
		Ackermann, and Weinzierl}]{aspock2020benchmark}
	\bibinfo{author}{L.~Asp{\"o}ck}, \bibinfo{author}{M.~Vorl{\"a}nder},
	\bibinfo{author}{F.~Brinkmann}, \bibinfo{author}{D.~Ackermann},
	\bibinfo{author}{S.~Weinzierl}, \bibinfo{title}{Benchmark for room acoustical
		simulation (bras)}, \bibinfo{year}{2020}. \URLprefix
	\url{10.14279/depositonce-6726.3}, \bibinfo{note}{from Audio Communication
		Group}.
	\bibitem[{Noxon(1987)}]{noxon1987controlled}
	\bibinfo{author}{A.~M. Noxon},
	\newblock \bibinfo{title}{Controlled reflection isolation booth},
	\newblock in: \bibinfo{booktitle}{Audio Engineering Society Convention 83},
	\bibinfo{organization}{Audio Engineering Society}, \bibinfo{year}{1987}.
	\bibitem[{Eyring(1930)}]{eyring1930reverberation}
	\bibinfo{author}{C.~F. Eyring},
	\newblock \bibinfo{title}{Reverberation time in “dead” rooms},
	\newblock \bibinfo{journal}{The Journal of the Acoustical Society of America}
	\bibinfo{volume}{1} (\bibinfo{year}{1930}) \bibinfo{pages}{217--241}.
	\bibitem[{Xiang(2020)}]{xiang2020generalization}
	\bibinfo{author}{N.~Xiang},
	\newblock \bibinfo{title}{Generalization of sabine's reverberation theory},
	\newblock \bibinfo{journal}{The Journal of the Acoustical Society of America}
	\bibinfo{volume}{148} (\bibinfo{year}{2020}) \bibinfo{pages}{R5--R6}.
	\bibitem[{Eyring(1931)}]{eyring1931reverberation}
	\bibinfo{author}{C.~F. Eyring},
	\newblock \bibinfo{title}{Reverberation time measurements in coupled rooms},
	\newblock \bibinfo{journal}{The Journal of the Acoustical Society of America}
	\bibinfo{volume}{3} (\bibinfo{year}{1931}) \bibinfo{pages}{181--206}.
	\bibitem[{Schroeder(1965)}]{schroeder1965new}
	\bibinfo{author}{M.~R. Schroeder},
	\newblock \bibinfo{title}{New method of measuring reverberation time},
	\newblock \bibinfo{journal}{The Journal of the Acoustical Society of America}
	\bibinfo{volume}{37} (\bibinfo{year}{1965}) \bibinfo{pages}{1187--1188}.
	\bibitem[{Billon et~al.(2006)Billon, Valeau, Sakout, and
		Picaut}]{billon2006use}
	\bibinfo{author}{A.~Billon}, \bibinfo{author}{V.~Valeau},
	\bibinfo{author}{A.~Sakout}, \bibinfo{author}{J.~Picaut},
	\newblock \bibinfo{title}{On the use of a diffusion model for acoustically
		coupled rooms},
	\newblock \bibinfo{journal}{The Journal of the Acoustical Society of America}
	\bibinfo{volume}{120} (\bibinfo{year}{2006}) \bibinfo{pages}{2043--2054}.
	\bibitem[{Prawda et~al.(2022)Prawda, Schlecht, and
		V{\"a}lim{\"a}ki}]{prawda2022calibrating}
	\bibinfo{author}{K.~Prawda}, \bibinfo{author}{S.~J. Schlecht},
	\bibinfo{author}{V.~V{\"a}lim{\"a}ki},
	\newblock \bibinfo{title}{Calibrating the sabine and eyring formulas},
	\newblock \bibinfo{journal}{The Journal of the Acoustical Society of America}
	\bibinfo{volume}{152} (\bibinfo{year}{2022}) \bibinfo{pages}{1158--1169}.
	\bibitem[{Long(2014)}]{LONG2014829}
	\bibinfo{author}{M.~Long},
	\newblock \bibinfo{title}{Design of studios and listening rooms},
	\newblock in: \bibinfo{editor}{M.~Long} (Ed.),
	\bibinfo{booktitle}{Architectural Acoustics}, \bibinfo{edition}{second
		edition} ed., \bibinfo{publisher}{Academic Press}, \bibinfo{address}{Boston},
	\bibinfo{year}{2014}, pp. \bibinfo{pages}{829--871}.
	\bibitem[{Mei and Kang(2012)}]{mei2012experimental}
	\bibinfo{author}{H.~Mei}, \bibinfo{author}{J.~Kang},
	\newblock \bibinfo{title}{An experimental study of the sound field in a large
		atrium},
	\newblock \bibinfo{journal}{Building and environment} \bibinfo{volume}{58}
	(\bibinfo{year}{2012}) \bibinfo{pages}{91--102}.
	\bibitem[{D'Orazio et~al.(2020)D'Orazio, Montoschi, and Garai}]{d2020acoustic}
	\bibinfo{author}{D.~D'Orazio}, \bibinfo{author}{F.~Montoschi},
	\bibinfo{author}{M.~Garai},
	\newblock \bibinfo{title}{Acoustic comfort in highly attended museums: A
		dynamical model},
	\newblock \bibinfo{journal}{Building and Environment} \bibinfo{volume}{183}
	(\bibinfo{year}{2020}) \bibinfo{pages}{107176}.
	\bibitem[{Prato et~al.(2016)Prato, Casassa, and
		Schiavi}]{prato2016reverberation}
	\bibinfo{author}{A.~Prato}, \bibinfo{author}{F.~Casassa},
	\bibinfo{author}{A.~Schiavi},
	\newblock \bibinfo{title}{Reverberation time measurements in non-diffuse
		acoustic field by the modal reverberation time},
	\newblock \bibinfo{journal}{Applied Acoustics} \bibinfo{volume}{110}
	(\bibinfo{year}{2016}) \bibinfo{pages}{160--169}.
	\bibitem[{Jackson(2021)}]{jackson2021classical}
	\bibinfo{author}{J.~D. Jackson}, \bibinfo{title}{Classical electrodynamics},
	\bibinfo{publisher}{John Wiley \& Sons}, \bibinfo{year}{2021}.
	\bibitem[{Johnson et~al.(1987)Johnson, Koplik, and Dashen}]{johnson1987theory}
	\bibinfo{author}{D.~L. Johnson}, \bibinfo{author}{J.~Koplik},
	\bibinfo{author}{R.~Dashen},
	\newblock \bibinfo{title}{Theory of dynamic permeability and tortuosity in
		fluid-saturated porous media},
	\newblock \bibinfo{journal}{Journal of fluid mechanics} \bibinfo{volume}{176}
	(\bibinfo{year}{1987}) \bibinfo{pages}{379--402}.
	\bibitem[{Champoux and Allard(1991)}]{champoux1991dynamic}
	\bibinfo{author}{Y.~Champoux}, \bibinfo{author}{J.-F. Allard},
	\newblock \bibinfo{title}{Dynamic tortuosity and bulk modulus in air-saturated
		porous media},
	\newblock \bibinfo{journal}{Journal of applied physics} \bibinfo{volume}{70}
	(\bibinfo{year}{1991}) \bibinfo{pages}{1975--1979}.
	\bibitem[{Maa(1998)}]{maa1998potential}
	\bibinfo{author}{D.-Y. Maa},
	\newblock \bibinfo{title}{Potential of microperforated panel absorber},
	\newblock \bibinfo{journal}{the Journal of the Acoustical Society of America}
	\bibinfo{volume}{104} (\bibinfo{year}{1998}) \bibinfo{pages}{2861--2866}.
	\bibitem[{Qu(3412)}]{qu2022broadband}
	\bibinfo{author}{S.~Qu}, \bibinfo{title}{Broadband Microwave and Underwater
		Acoustic Absorption Metamaterials: Approaching the Causal Limit}, Ph.D.
	thesis, The Hong Kong University of Science and Technology,
	\bibinfo{year}{2022}.
	\URLprefix \url{https://doi.org/10.14711/thesis-991013100058903412}.
	\bibitem[{Qu et~al.(2021)Qu, Hou, and Sheng}]{qu2021conceptual}
	\bibinfo{author}{S.~Qu}, \bibinfo{author}{Y.~Hou}, \bibinfo{author}{P.~Sheng},
	\newblock \bibinfo{title}{Conceptual-based design of an ultrabroadband
		microwave metamaterial absorber},
	\newblock \bibinfo{journal}{Proceedings of the National Academy of Sciences}
	\bibinfo{volume}{118} (\bibinfo{year}{2021}) \bibinfo{pages}{e2110490118}.
	\bibitem[{Qu et~al.(2022)Qu, Gao, Tinel, Morvan, Romero-Garc{\'\i}a, Groby, and
		Sheng}]{qu2022underwater}
	\bibinfo{author}{S.~Qu}, \bibinfo{author}{N.~Gao}, \bibinfo{author}{A.~Tinel},
	\bibinfo{author}{B.~Morvan}, \bibinfo{author}{V.~Romero-Garc{\'\i}a},
	\bibinfo{author}{J.-P. Groby}, \bibinfo{author}{P.~Sheng},
	\newblock \bibinfo{title}{Underwater metamaterial absorber with
		impedance-matched composite},
	\newblock \bibinfo{journal}{Science Advances} \bibinfo{volume}{8}
	(\bibinfo{year}{2022}) \bibinfo{pages}{eabm4206}.
	\bibitem[{ISO 3382-1:2008; 3382-2:2008(2008)}]{ISO}
	ISO 3382-1:2008; 3382-2:2008, \bibinfo{title}{{Acoustics — Measurement of
			room acoustic parameters — Part 1: Performance spaces; Part 2:
			Reverberation time in ordinary rooms}}, \bibinfo{type}{Standard},
	International Organization for Standardization, \bibinfo{address}{Geneva,
		CH}, \bibinfo{year}{2008}.
	\bibitem[{Barron(2009)}]{barron2009auditorium}
	\bibinfo{author}{M.~Barron}, \bibinfo{title}{Auditorium acoustics and
		architectural design}, \bibinfo{publisher}{Routledge}, \bibinfo{year}{2009}.
	\bibitem[{Cecchi et~al.(2017)Cecchi, Carini, and Spors}]{cecchi2017room}
	\bibinfo{author}{S.~Cecchi}, \bibinfo{author}{A.~Carini},
	\bibinfo{author}{S.~Spors},
	\newblock \bibinfo{title}{Room response equalization—a review},
	\newblock \bibinfo{journal}{Applied Sciences} \bibinfo{volume}{8}
	(\bibinfo{year}{2017}) \bibinfo{pages}{16}.
\end{thebibliography}


\end{document}